\documentclass[aps,prb,twocolumn,showpacs,psfig,superscriptaddress,longbibliography]{revtex4-1}
\usepackage{textcomp}
\usepackage{times}
\usepackage{graphicx}
\usepackage{float}
\usepackage{latexsym,amsmath,amssymb,bm,euscript}
\usepackage{color}
\usepackage{subfigure}
\usepackage{epstopdf}
\usepackage[colorlinks=true,linkcolor=blue,citecolor=blue]{hyperref}
\usepackage{hyperref}
\usepackage{soul}
\usepackage[normalem]{ulem}
\usepackage{mathrsfs}
\usepackage{amsmath}
\usepackage{lettrine}
\usepackage{xspace}
\usepackage{textcomp}
\usepackage{xr}
\usepackage{appendix}
\usepackage{verbatim}

\begin{document}

\title{NMR evidence of spinon localization in kagome antiferromagnet YCu$_3$(OH)$_6$Br$_2$[Br$_{1-x}$(OH)$_x$]}

\author{Shuo Li}
\thanks{These authors contributed equally to this study.}
\affiliation{Department of Physics and Beijing Key Laboratory of
Opto-electronic Functional Materials $\&$ Micro-nano Devices, Renmin
University of China, Beijing, 100872, China}

\author{Yi Cui}
\thanks{These authors contributed equally to this study.}
\affiliation{Department of Physics and Beijing Key Laboratory of
Opto-electronic Functional Materials $\&$ Micro-nano Devices, Renmin
University of China, Beijing, 100872, China}
\affiliation{Key Laboratory of Quantum State Construction and Manipulation (Ministry of Education),
Renmin University of China, Beijing, 100872, China}

\author{Zhenyuan Zeng}
\thanks{These authors contributed equally to this study.}
\affiliation{Beijing National Laboratory for Condensed Matter Physics and Institute of Physics,
Chinese Academy of Sciences, Beijing, 100190, China}

\author{Yue Wang}
\thanks{These authors contributed equally to this study.}
\affiliation{Department of Physics and Beijing Key Laboratory of
Opto-electronic Functional Materials $\&$ Micro-nano Devices, Renmin
University of China, Beijing, 100872, China}

\author{Ze Hu}
\affiliation{Department of Physics and Beijing Key Laboratory of
Opto-electronic Functional Materials $\&$ Micro-nano Devices, Renmin
University of China, Beijing, 100872, China}

\author{Jie Liu}
\affiliation{Department of Physics and Beijing Key Laboratory of
Opto-electronic Functional Materials $\&$ Micro-nano Devices, Renmin
University of China, Beijing, 100872, China}

\author{Cong Li}
\affiliation{Department of Physics and Beijing Key Laboratory of
Opto-electronic Functional Materials $\&$ Micro-nano Devices, Renmin
University of China, Beijing, 100872, China}

\author{Xiaoyu Xu}
\affiliation{Department of Physics and Beijing Key Laboratory of
Opto-electronic Functional Materials $\&$ Micro-nano Devices, Renmin
University of China, Beijing, 100872, China}

\author{Ying Chen}
\affiliation{Department of Physics and Beijing Key Laboratory of
Opto-electronic Functional Materials $\&$ Micro-nano Devices, Renmin
University of China, Beijing, 100872, China}


\author{Zhengxin Liu}
\email{liuzxphys@ruc.edu.cn}
\affiliation{Department of Physics and Beijing Key Laboratory of
Opto-electronic Functional Materials $\&$ Micro-nano Devices, Renmin
University of China, Beijing, 100872, China}
\affiliation{Key Laboratory of Quantum State Construction and Manipulation (Ministry of Education),
Renmin University of China, Beijing, 100872, China}

\author{Shiliang Li}
\email{slli@iphy.ac.cn}
\affiliation{Beijing National Laboratory for Condensed Matter Physics and
Institute of Physics, Chinese Academy of Sciences, Beijing, 100190, China}
\affiliation{School of Physical Sciences, University of Chinese Academy of Sciences, Beijing, 100190, China}
\affiliation{Songshan Lake Materials Laboratory, Dongguan, Guangdong, 523808, China}

\author{Weiqiang Yu}
\email{wqyu\_phy@ruc.edu.cn}
\affiliation{Department of Physics and Beijing Key Laboratory of
Opto-electronic Functional Materials $\&$ Micro-nano Devices, Renmin
University of China, Beijing, 100872, China}
\affiliation{Key Laboratory of Quantum State Construction and Manipulation (Ministry of Education),
Renmin University of China, Beijing, 100872, China}


\begin{abstract}
We performed nuclear magnetic resonance studies on a kagome antiferromagnet YCu$_3$(OH)$_6$Br$_2$[Br$_{1-x}$(OH)$_{x}$].
No significant NMR spectral broadening is found in the Br center peak from 1~K down to 0.05~K,
 indicating absence of static antiferromagnetic ordering.
In contrast to signatures of dominant 2D kagome antiferromagnetic fluctuations at temperature above 30~K, both the Knight shift $K_{\rm{n}}$ and the spin-lattice relaxation rate $1/T_{1}$ increase when the sample is cooled from 30~K to 8~K, which can be attributed to the scattering of spin excitations by strong non-magnetic impurities.
Unusually, a hump is observed  in $K_{\rm{n}}$ and $1/T_{2}$ close to 2~K (far below the exchange energy), which indicates the existence of excitations with a large density of states close to zero energy.
These phenomena are reproduced by a mean-field simulation of Heisenberg model with bond-dependent exchange interactions,
where the sign fluctuations in the spinon kinetic terms caused by impurities 
result in localization of spinons and an almost flat band close to the Fermi energy.
\end{abstract}

\maketitle


\section{\label{Intro} Introduction}

Quantum spin liquids (QSLs) are an exotic phase of matter characterized by long-range quantum entanglement of spins with absence of long range magnetic order~\cite{2010_Nature_Balents,2017_RMP_YiZhou,2020_Science_Senthil}.
Generally a QSL does not break any symmetry of the spin Hamiltonian and hosts fractional spinon excitations~\cite{1987_PRB_Sethna,1989_PRB_Chakraborty,1991_PRB_XGWen}.
It was proposed that geometrical frustration in antiferromagnets can effectively enhance quantum fluctuations and suppress magnetic order. Therefore, QSL candidates have been extensively studied in antiferromagnets with triangular~\cite{2003_PRL_Saito,2014_PRL_Hatsumi,2016_PRL_Satio,2017_PRL_YSLi}, kagome~\cite{2012_Nature_YSLee,2014_PRL_Schlueter,2015_Science_YSLee,2016_RMP_Norman,2018_PRM_Krellner}, and pyrochlore~\cite{1997_PRL_Godfrey,2001_science_Gingras} lattices.
Especially, the strongly frustrated spin-1/2 kagome Heisenberg antiferromagnet (KHAF) is considered as an ideal model to stabilize QSL ground states~\cite{2015_Science_YSLee}.
Theoretical works reveal that KHAF can exhibit no magnetic order even at zero temperature, but whether the ground state is gapped~\cite{2011_Science_White,2012_PRL_Schollwock} or  gapless~\cite{2007_PRL_XGWen,2008_PRB_XGWen,2011_PRB_Poilblanc,2017_PRX_Pollmann,2018_SciBull_GSu} remains controversial.

However, most of magnetic materials in nature exhibit long-range magnetic ordering at low temperatures, which makes it challenging to study QSL phases.
On experimental side, the mineral herbertsmithite ZnCu$_3$(OH)$_6$Cl$_2$ was thought to realize the
ideal spin-$1/2$ KHAF model~\cite{2005_JACS_Nocera,2007_PRL_YSLee,2008_PRL_Harrison,2011_PRB_YSLee,2012_Nature_YSLee,2015_Science_YSLee,2021_NaturePhy_Imai},
in which Cu$^{2+}$ ions with spin-$1/2$ moments arrange in structurally perfect kagome layers and non-magnetic Zn$^{2+}$ ions separate lattice planes to avoid 3D long-range antiferromagnetic (AFM) order.
Accordingly, experiments on these materials are expected to provide a platform for determining spin excitations~\cite{2012_Nature_YSLee,2015_Science_YSLee}.
However, antisite randomness, that about 15\% Cu$^{2+}$ impurities occupy the non-magnetic Zn$^{2+}$ sites within the interlayers outside the kagome planes and 1\% Zn$^{2+}$ defects occupy the Cu$^{2+}$ sites within the kagome planes, strongly affects the low-energy probes~\cite{2008_PRL_Harrison,2010_JACS_Nocera,2011_PRB_YSLee,2011_PRL_Harrison,2012_Nature_YSLee,2015_Science_YSLee,2020_NaturePhy_Mendels,2021_NaturePhy_Imai}.
Bulk susceptibility $\chi$ is strongly enhanced by interlayer magnetic defects, preventing accurate measurements of intrinsic properties of the kagome layers~\cite{2008_PRL_Harrison,2011_PRB_YSLee}.
It was further proposed that antisite randomness could  disrupt QSL ground states~\cite{2021_PRL_SYLi}.
Recent studies show that certain fraction of Cu$^{2+}$ spins form robust QSL-like spin-singlets with spatially varying gaps~\cite{2021_NaturePhy_Imai,2021_PRL_SYLi,2022_PRL_Takashi}.
However, the fraction of spin-singlets does not exceed 60\% at $T{\sim}0.01 J$ and other Cu$^{2+}$ spins remain paramagnetic~\cite{2021_NaturePhy_Imai}.
Alternatively, random singlet was also proposed as a candidate state in the 2D KHAF due to antisite randomness~\cite{1994_PRB_Fisher,2010_PRL_Singh}.

To avoid magnetic antisite disorder, Y-kapellasite  YCu$_3$(OH)$_{6+x}$Cl$_{3-x}$ with $x$~$=$~0, 1/3 have been synthesized.
For $x$~$=$~0, the compound exhibits AFM order at $T_N$~$\approx$~15~K\cite{2016_JMCC_MJX,2019_PRM_Fabrice,2020_PRL_Zorko,2022_arXiv_Pustogow}, which is partly attributed to a strong Dzyaloshinskii-Moriya (DM) interaction\cite{2022_npj_Johannes,2023_PRB_Bert}.
For $x$~$=$~1/3, namely Y$_3$Cu$_9$(OH)$_{19}$Cl$_8$, with distorted kagome lattice of Cu$^{2+}$ which yields three different nearest-neighbor interactions, has a $T_N$~$\approx$~2.2~K~\cite{2017_JMCC_Krellner,2022_arXiv_Pustogow, 2023_PRB_Bert}.

Recently, YCu$_3$(OH)$_6$Br$_2$ [Br$_{1-x}$(OH)$_{x}$] (YCu$_3$-Br) was also synthesized and reported as a QSL candidate
with no magnetic ordering down to 2~K~\cite{2020_JMMM_JXM}. The material has perfect Cu$^{2+}$ kagome planes with nearest exchange couplings $J~{\approx}$~79~K~\cite{2020_JMMM_JXM,2022_PRB_YSLi,2022_PRB_SLLi2,2022_Commphy_YSLi,2022_PRB_Hess}.
However, non-magnetic antisite randomness among interlayer Br$^-$ and OH$^-$ exists because of their
similar sizes~\cite{2020_JMMM_JXM,2022_PRB_YSLi,2022_PRB_SLLi2,2022_Commphy_YSLi,2022_PRB_Hess}.
In fact, recent studies indicate that substitution of OH$^{-}$ for Br$^-$ results in a non-symmetrical Cu$^{2+}$ hexagon
in the kagome lattices, which leads to local distortions in the Cu-O-Cu exchange paths and
therefore bond-dependent, non-uniform exchange couplings~\cite{2022_Commphy_YSLi}.
The impact of this type of antisite disorder to the ground state needs to be studied.

In this paper, we report $^{79}$Br and $^{81}$Br nuclear magnetic resonance (NMR) measurements on assembled high-quality single crystals of YCu$_3$-Br ($x$~$=$~0.67).
We found that the full width at half maximum (FWHM) of the NMR spectra barely change with temperature from 1~K down to 50~mK. Combined with
absence of magnetic ordering at higher temperatures by other studies, no static magnetic ordering in the system is concluded.
Upon cooling to 30~K, the system is dominated by 2D KHAF physics as revealed by moderate increasing of the Knight shift $K_{\rm{n}}$ towarding a peaked behavior.
Below 30~K, Knight shift $K_{\rm{n}}$, spin-lattice relaxation rate $1/T_1$, and spin-spin relaxation rate $1/T_2$ all increase upon cooling but with additive anisotropic and isotropic components, which suggests enhanced low-energy spin fluctuations from both interlayer couplings and disorder effects.
A peak shows up below 2~K in both $K_{\rm{n}}$ and $1/T_2$, indicating that the spinons are scattered by non-magnetic impurities and form a large amount of 
magnetic excitations with small but nonzero 
energy.

\begin{figure}[t]
\includegraphics[width=8.7cm]{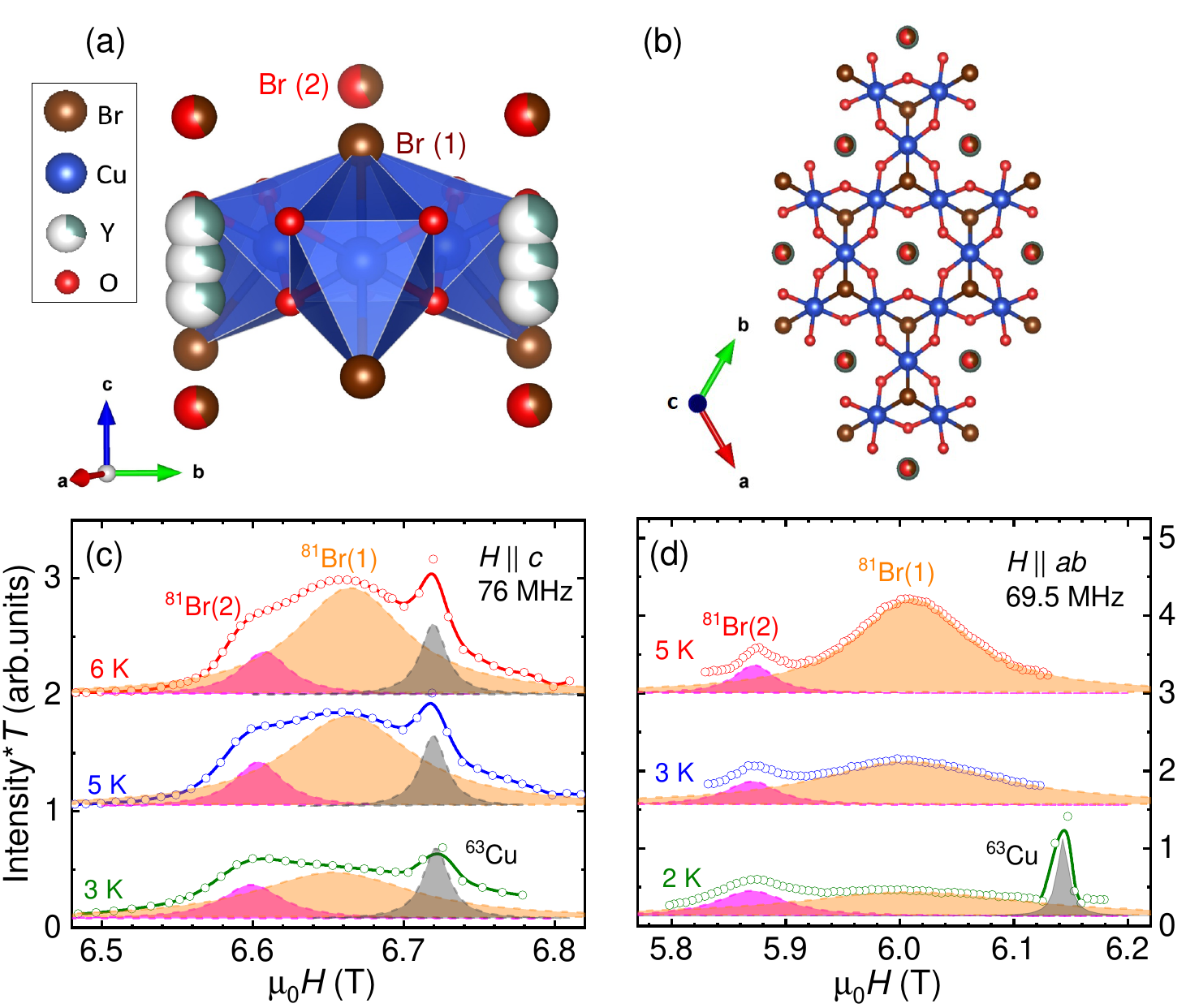}
\caption{\label{struc1}
\textbf{Crystal structure and presentative spectra of YCu$_3$-Br.}
(a) Side view of the crystal structure, in which hydrogens are not shown.
(b) Top view of the kagome plane.
Cu$^{2+}$ atoms are surrounded by O atoms and Br atoms.
Two inequivalent Br sites are shown, with Br(1) above and below the center of Cu$^{2+}$ triangles and Br(2) located at the interlayer which is partly occupied by OH$^{-}$.
Y atoms reside on three positions (cyan) with the occupancy ratio of 35$\%$, 30$\%$, and 35$\%$, respectively.
(c) Br spectra measured with $H||c$ at a fixed frequency 76~MHz.
$^{81}$Br(1) and $^{81}$Br(2) spectra partly overlap with $^{63}$Cu spectra (grey) from the NMR coil.
(d) Br spectra measured with $H||ab$ at a fixed frequency 69.5~MHz, also partially overlapping with environmental $^{63}$Cu spectra as labeled.
}
\end{figure}

\section{\label{Materials} Material and Techniques}

Single crystals of YCu$_3$-Br ($x$~$=$~0.67) were grown by hydrothermal method as reported previously~\cite{2022_PRB_SLLi2}.
The lattice structure of YCu$_3$-Br is shown in Fig.~\ref{struc1}(a) and (b), where Cu$^{2+}$ ions form 2D kagome planes separated by Br$^-$ layers. There are two inequivalent Br$^-$ sites in the lattice, labeled as Br(1) and Br(2) respectively.
The nonsymmetric distribution of OH$^{-}$/Br$^{-}$ pushes 70\% of Y$^{3+}$ away from the ideal position, resulting in Y$^{3+}$ occupancy probability to be 35$\%$,
30$\%$ and 35$\%$, from the top to the bottom, respectively.

For our NMR measurements, several crystals were aligned to a total size of 5$\times$6$\times$1.2~mm$^3$, in order to improve
the signal to noise ratio [see Sec.~S6 in supplementary materials (SM)~\cite{supply}]. The samples were cooled in a variable-temperature-insert for temperature above 2~K and a dilution refrigerator for temperature below 2~K.
Spectra of $^{79}$Br and $^{81}$Br isotopes, having nuclear spin $I$~$=$~$3/2$, Zeeman factor of $^{79}\gamma$~$=$~10.667 MHz/T and $^{81}\gamma$~$=$~11.499 MHz/T, and quadrupole moment of $^{79}Q$~$=$~0.33$\times$10$^{-28}$m$^2$ and $^{81}Q$~$=$~0.28$\times$10$^{-28}$m$^2$, respectively, were collected with the standard spin-echo sequences $\pi$/2-$\tau$-$\pi$ where time length of the $\pi/2$ pulse is about 2~$\mu s$.
The full NMR spectra are obtained by integrating spin-echo spectra with field sweeping at fixed frequencies.

\begin{figure}[t]
\includegraphics[width=8.6cm]{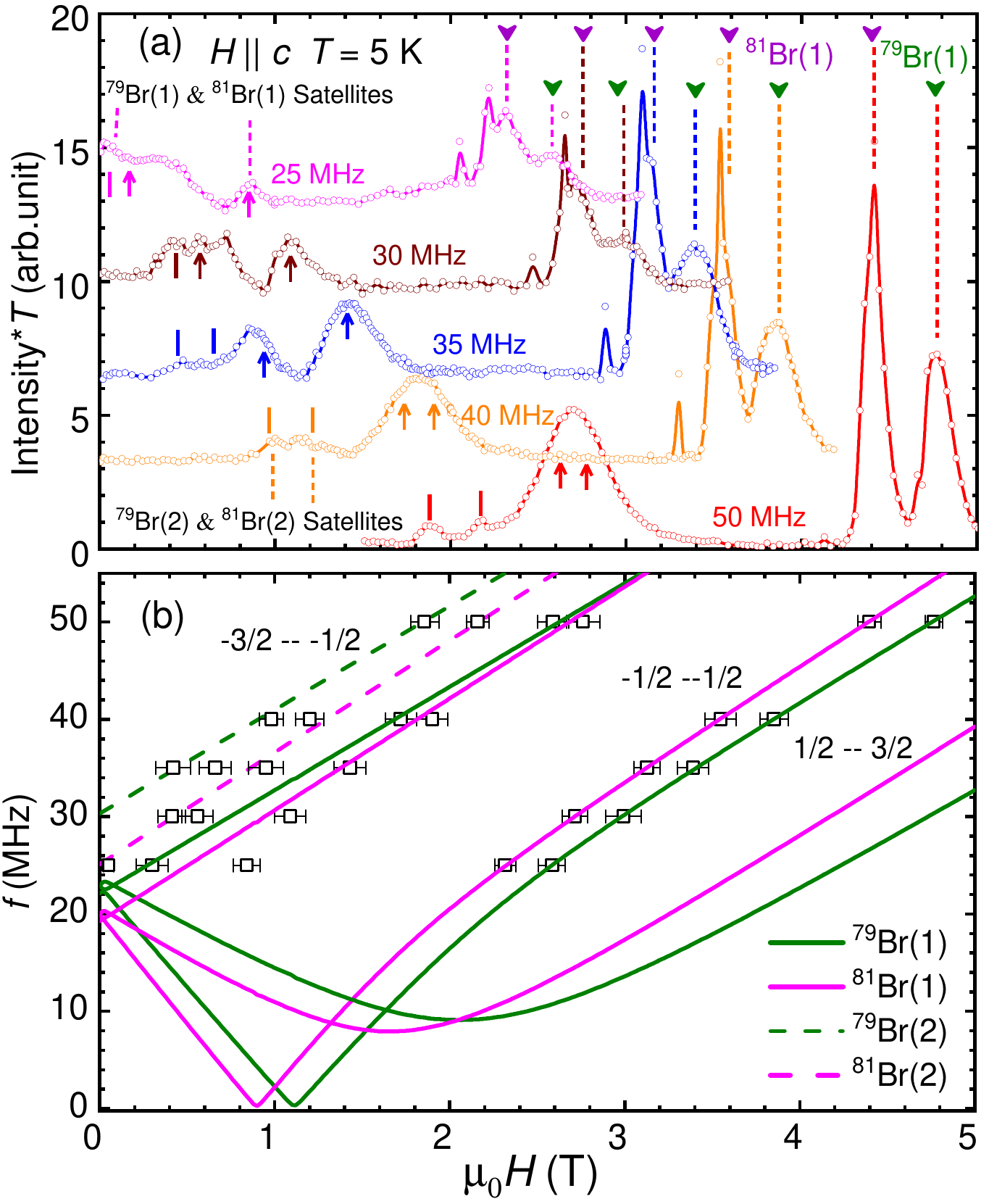}
\caption{\label{spec2}
\textbf{Br spectra at 5~K with $H||c$.}
(a) Br NMR spectra measured with fixed frequencies as functions of fields.
Vertical offsets are applied for clarity.
The downward arrows mark the locations of the Br(1) central peaks, and the upward arrows mark the Br(1) satellite peaks.
The vertical solid lines represent the satellites of Br(2).
(b) Peak frequencies of the spectra as functions of fields.
The solid lines are fit to the resonance frequencies,
from contributions of different nuclear sites and isotopes as labeled (see text).
Corresponding assignments of spectra to isotopes are also labeled in panel (a).
}
\end{figure}

\begin{figure*}[t]
\includegraphics[width=17cm]{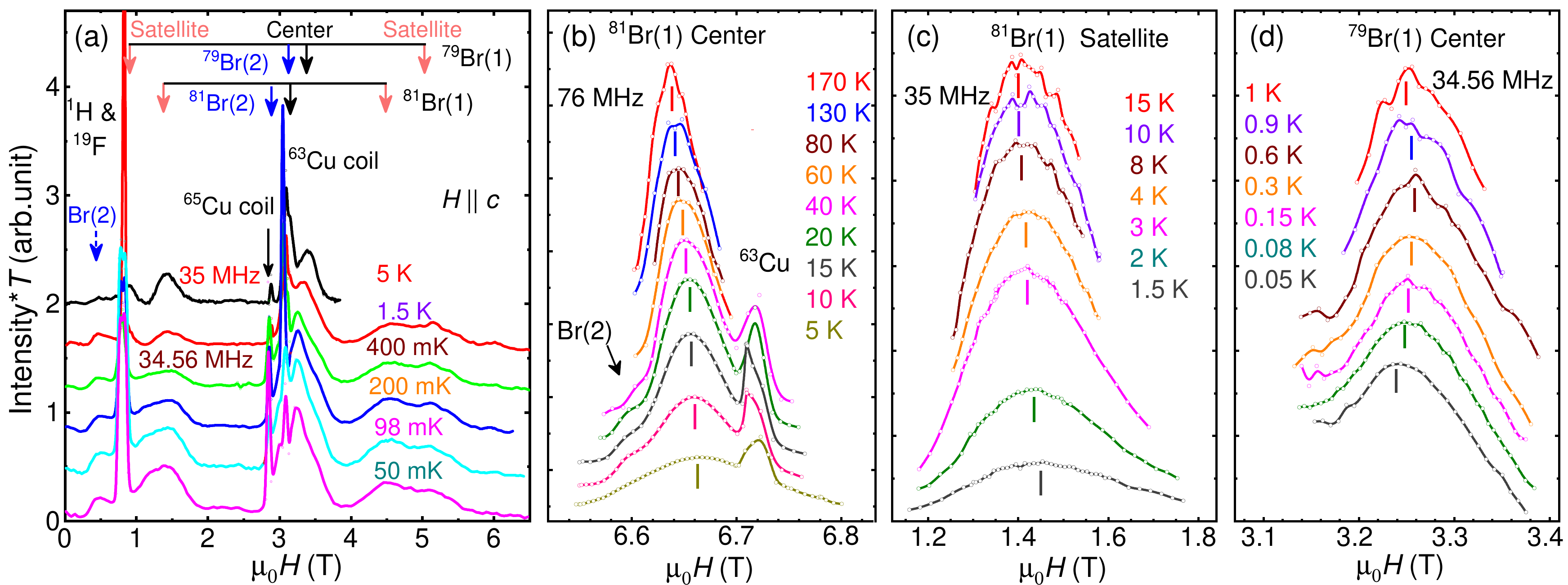}
\caption{\label{spec3}
\textbf{Br Spectra at different temperatures with $H||c$.}
(a) Presentative full spectra acquired as functions of fields,
with center and satellite transition lines of Br(1) and Br(2) specified.
$^{1}$H, $^{19}$F and $^{63}$Cu note the environmental NMR signals.
(b)-(d) $^{81}$Br(1) lines at high, intermediate, and low temperatures, respectively. Vertical lines mark the peak positions.
}
\end{figure*}

We primarily study the spin fluctuations on the Br(1) site, because the spectral weight of Br(2) is expected to be 1/6 of that of Br(1), following their occupancy ratios in the lattice.
The Knight shift $K_{\rm{n}}$ is calculated by the form $K_{\rm{n}}$~$=$~$(f-f_{\nu})/{\gamma}H-1$, where $f$ is peak frequency in the spectrum, $f_{\nu}$ is the frequency correction due to  quadrupole moment contributions, and $H$ is external field.
The spin-lattice relaxation rate $1/T_1$ was measured by inversion-recovery method, by fitting nuclear magnetization to the standard recovery function
for spin-3/2 isotopes, $M(t)$~$=$~$M_{0}[1-0.1e^{-(t/T_{1})^{\beta}}-0.9e^{-(6t/T_{1})^{\beta}}]$, where $\beta$ is stretching factor.
The spin-spin relaxation rate $1/T_{2}$ was obtained by fitting the transverse spin recovery with Lorentz form, that is, $M(t)$~$=$~$M_{0}[0.1e^{-(t/T_{2})^{\beta}}+0.9e^{-(6t/T_{2})^{\beta}}]$.
The detailed spin recovery curves and the fittings are presented in Sec.~S6 of the SM~\cite{supply}.

\section{\label{sspec}NMR spectra}

\begin{figure}[t]
\includegraphics[width=8.2cm]{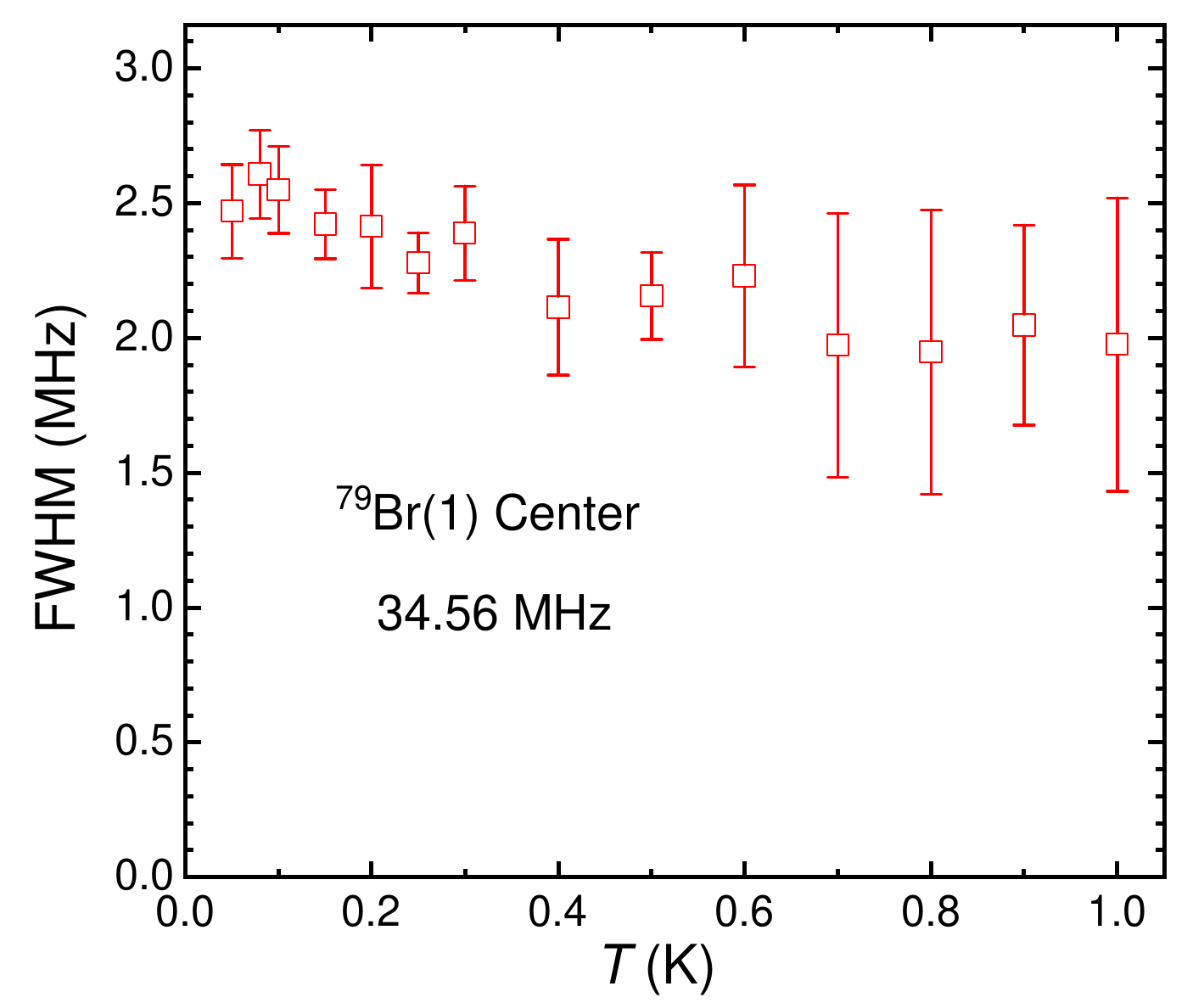}
\caption{\label{fwhm4}
\textbf{Low-temprature FWHM of the spectra.}
The FWHM with temperature from 1~K to 50~mK, with $H||c$, obtained from Fig.~\ref{spec3}(d).
}
\end{figure}

We performed $^{79}$Br and $^{81}$Br NMR measurements at different frequencies.
Typical spectra are partly demonstrated in Fig.~\ref{struc1}(c)-(d) and Fig.~\ref{spec2}(a), and full spectra in Fig.~\ref{spec3}(a), with different field orientations and field ranges.
In principle, twelve NMR lines are expected in each full spectrum, considering two inequivalent sites and two types of spin-$3/2$ Br isotopes.
Each type of isotope produces one center transition and two satellite transition lines in spectrum at a constant field, due to hyperfine field from Cu$^{2+}$, and coupling among the nuclear quadrupole moments and the local electric field gradient (EFG).

Due to antisite disorder, the change of O$^{2-}$ position in Cu-O-Cu bond results in the variation in both the EFG and the hyperfine field on
the Br sites, which broadens the Br NMR spectra by both a quadrupolar and a magnetic effect as observed in this study.

The frequencies of observed resonance peaks are then plotted in Fig.~\ref{spec2}(b) as functions of fields, which can be assigned to different sites and isotopes, following three conditions.
First, the relative spectral weight of Br(2) and Br(1) should follow a ratio of 1:6 as described in Sec.~\ref{Materials}. Second, the hyperfine shift of Br(2) should be
smaller compared to that of Br(1), because of its larger distance to magnetic Cu$^{2+}$ ions.
Third, $^{79}$Br has a smaller Zeeman factor and a larger quadrupole moment compared to  $^{81}$Br, which results in a smaller resonance frequency for $^{79}$Br center peaks but a larger quadrupole correction for satellite peaks.
By reading the frequency at the spectral of NMR and NQR spectra,
the peak frequencies of the spectrum can be fit with parameters $^{79}\nu_{Q}(1)$~$=$~24.3~MHz, $^{81}\nu_{Q}(1)$~$=$~21.5~MHz, $^{79}\nu_{Q}(2)$~$=$~30.3~MHz
and $^{81}\nu_{Q}(2)$~$=$~25.09~MHz, which are demonstrated by the fit curves of resonance frequencies in Fig.~\ref{spec2}(b).
The detailed fitting procedure is included in the SM~\cite{supply}.
The fit curves of the solid and dotted lines are in good consistency with the experimental data.
Then all the resonance peaks are assigned to different Br sites and isotopes as labeled in Fig.~\ref{struc1}(c)-(d) and Fig.~\ref{spec2}(a).

Figure~\ref{spec3}(a) presents full Br spectra from 5~K to 50~mK.
To reduce spectral overlaps among different transition lines, different parts of the Br(1) spectra are then measured with selected frequencies at three temperature ranges, that is, from 5~K to 170~K on center transitions (Fig.~\ref{spec3}(b)), from 1.5~K to 15~K on satellites (Fig.~\ref{spec3}(c)), and from 50~mK to 1~K on center transitions (Fig.~\ref{spec3}(d)).
The FWHM of the $^{79}$Br(1) spectra (Fig.~\ref{spec3}(d)) is obtained by Lorentz fitting and shown in Fig.~\ref{fwhm4}.

Upon cooling, the NMR spectra show a progressive broadening with temperatures from 8 K to 1.5 K.
However, no further line broadening is observed with temperatures from 1~K down to 50~mK,
and the NMR linewidth tends to saturate below 1~K (Fig.~\ref{fwhm4}).
In a previous report, the neutron scattering data exhibit no long-range magnetic order down to 0.3~K~\cite{2023_arxiv_SLLi}.
Combining these two pieces of information, our data provide evidence for absence of static antiferromagnetic ordering in YCu$_3$-Br with
temperature down to 50~mK.

\section{\label{skn}NMR Knight shift}

\begin{figure}[t]
\includegraphics[width=8.5cm]{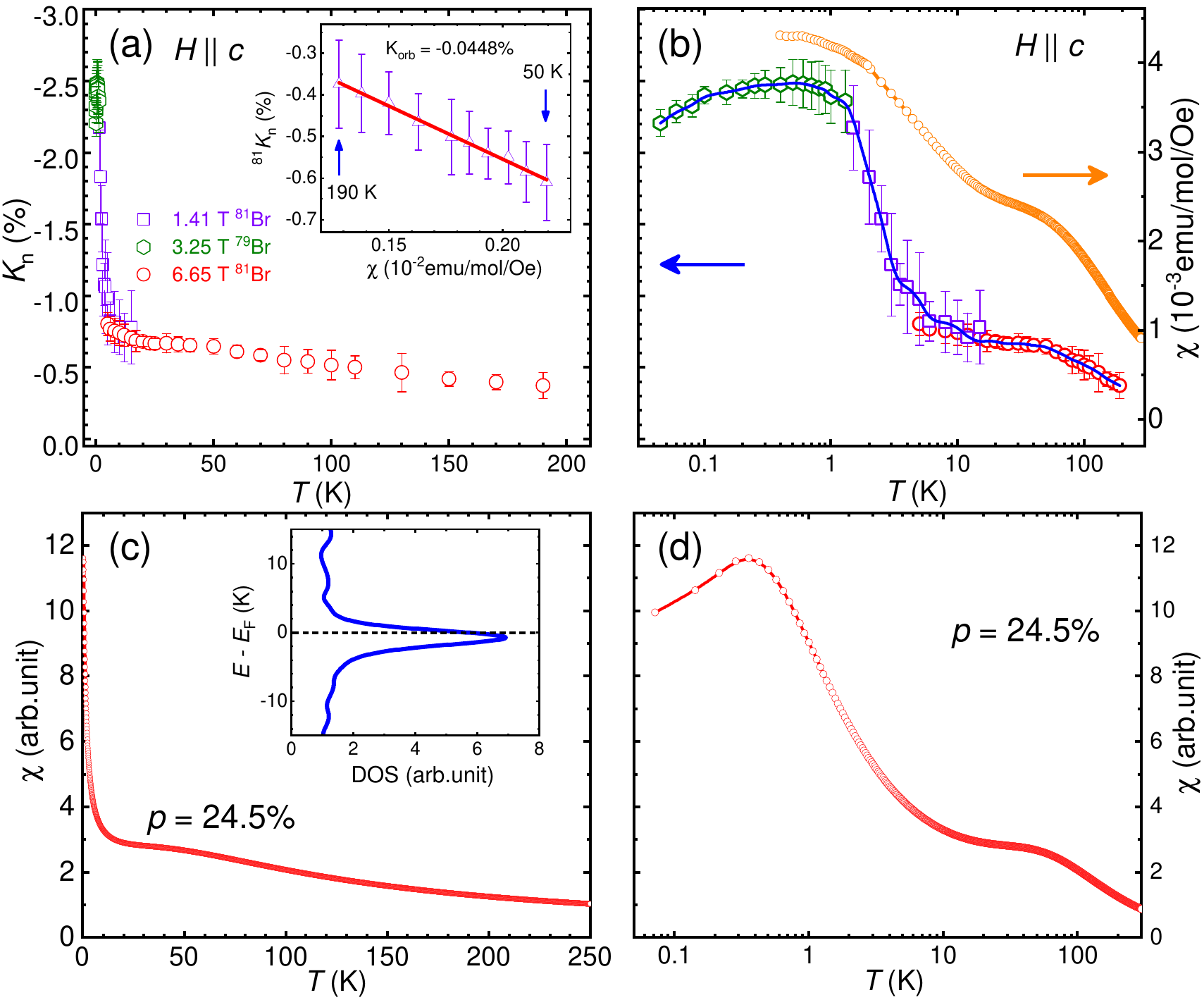}
\caption{\label{kn4}
\textbf{Experimental Knight shift, and theoretical bulk susceptibility by a mean-field calculation.}
(a) $K_{\rm{n}}$ as functions of temperatures measured with different fields.
Inset: $^{81}K_{\rm{n}}$ vs $\chi$ with temperatures from 50~K to 190~K.
(b) $K_{\rm{n}}$ (left) and $\chi$ (right) adapted from (Ref.~\citenum{2022_PRB_SLLi2}) of the crystals in a semilog scale.
(c) $\chi$ as a function of temperatures, calculated based on QSL state
with spinon Fermi-surface and bond disorder $p$~$=$~24.5$\%$ (see text).
Inset: Density of states as a function of energy.
(d) Semilog plot of $\chi$ as a function of temperatures.
}
\end{figure}

To reveal spin fluctuations of the system, the Knight shift $K_{\rm{n}}$, deduced from the Br(1) spectra shown in Fig.~\ref{spec3}(b)-(d), is plotted as functions of temperatures in Fig.~\ref{kn4}(a).
Note that $K_{\rm{n}}$ is determined by the frequency at the peak position of the spectra, and the error bar is calculated
by $\Delta f/\gamma H$, where $\Delta f$ is the frequency difference between the actual peak and the fitted peak (data not shown) in the spectra.
Upon cooling below 200~K, $K_{\rm{n}}$ first exhibits a slow increase with a shoulder behavior between the 30~K and 15~K, then undergoes a rapid raise from 8~K to 0.8~K, and finally decreases after reaching a peak at about 0.5~K.
The temperature dependence of $K_{\rm{n}}(T)$ is consistent with the bulk susceptibility $\chi(T)$ (adapted data from Ref.~\citenum{2022_PRB_SLLi2}) down to 0.5~K, as shown in Fig.~\ref{kn4}(b).
Together with the extremely low-temperature data, $K_{\rm{n}}$ reveals important information of the spin excitations at different energy scales~\cite{2008_PRL_Harrison, 2015_Science_YSLee, 2021_NaturePhy_Imai}.


Generally, $K_{\rm{n}}$~$=$~$K_{\rm s}+K_{\rm {orb}}$, where $K_{\rm {orb}}$ is the orbital contribution to Knight shift which does not change with temperature, and $K_{\rm s} (T) $ is the spin contribution.
Note that $K_{\rm s}(T)$~$=$~$A_{\rm hf}\chi(T)/N_A\mu_B$ , where $\chi(T)$ is the bulk spin susceptibility, $A_{\rm hf}$ is the magnetic hyperfine
coupling constant, and $N_A$ is the Avogadro's number.
The high-temperature $K_{\rm{n}}(T)$ is then plotted against $\chi(T)$ as shown in the inset of Fig.~\ref{kn4}(a), where they
follow a straight line and give the hyperfine coupling $A_{\rm hf}$~$=$~-4.74 $\pm$0.14 ~kOe$/\mu_{\rm B}$.
$K_{\rm orb}$ of -0.0448$\%$ is determined by the intercept with the $y$ axis at $\chi=0$.

It was proposed theoretically that $\chi(T)$ in the 2D KHAF yields a peak at temperature in the order of $T \sim 0.15J$~\cite{2018_SciBull_GSu}.
The shoulder behavior in both $K_n(T)$ and $\chi(T)$ between 15~K and 30~K indicates that the low-energy spin dynamics above 30~K
in this system is dominated by the intrinsic behavior of the 2D KHAF, given that $J\approx$~79~K~\cite{2022_PRB_YSLi,2022_PRB_SLLi2}.

Below 15~K, however, the absence of the rapid drop in $K_{\rm{n}}(T)$ and $\chi(T)$ indicates that another mechanism exists.
Indeed, $K_{\rm{n}}$ shows a remarkable upturn with decreasing temperature and reaches a peak at around 0.5~K.
At first glance, the peak at 0.5~K may be caused by a criticality toward 3D ordering owing to 3D interlayer coupling.
However, with such a high onset temperature of upturn, a very strong interlayer coupling has to be considered, which is contradictory with the absence of 3D long-range ordering as reported in Sec.~\ref{sspec}.
An alternative interpretation for the peak is that there exists a large density of 
low-energy magnetic excitations caused by non-magnetic impurities as discussed below.

We preformed theoretical simulations of 2D kagome Heisenberg model with bond-dependent interactions using fermionic spinon mean-field theory.
Considering that the ground state has no magnetic order, the spinon representation, with an unbroken rotational symmetry,
is convenient to describe the magnetic disorder ground state in the mean-field approach.
Under a local constraint $\sum_{\sigma}f_{i\sigma}^\dag f_{i\sigma}=1$, the exchange interaction is rewritten as $J_{ij}\pmb S_i\cdot\pmb S_j =  -\frac {J_{ij}}{2}\sum_{\alpha\beta}f_{i\alpha}^{\dagger}f_{j\alpha}f_{j\beta}^{\dagger}f_{i\beta}$.
Under mean field approximation, the spin-spin interactions are decoupled into the non-interacting trial Hamiltonian $H_{\rm MF} ~=~ \sum_{ij,\sigma}\eta_{ij}  (t_{ij} + \delta_{ij}) f_{i\sigma}^{+}f_{j\sigma}+ {\rm h.c.} + \lambda \sum_i f^\dag_{i\sigma}f_{i\sigma}$, where $f_{\uparrow}$ and $f_\downarrow$ are fermionic spinon operators, $\sigma$ is the spin index, $t_{ij}$ is the averaged hopping amplitude (the kinetic term) for the spinons, $\delta_{ij}$ is amplitude fluctuations in the hopping, $\eta_{ij}$ represents phase randomness in the hopping, and $\lambda$ is the Lagrangian multiplier for the constraint $\sum_\sigma f_{i\sigma}^\dag f_{i\sigma}$~$=$~$1$.
We emphasize that non-uniform exchange interactions $J_{ij}$ are the physical origin of
phase fluctuations $\eta_{ij}$ and amplitude fluctuations $\delta_{ij}$ in the spinon
kinetic energy term.


In the simulations, we firstly assume that the amplitude fluctuations $\delta_{ij}$ obey Gaussian distribution.
However, the amplitude fluctuations only moderately change the ground state (see SM~\cite{supply}).
Then we turn on the phase fluctuations $\eta_{ij}$ and simply consider them as random sign distributions due to the unbroken time reversal symmetry, namely, $\eta_{ij}$~$=$~${\pm}1$ with probability $P(-1)$~$=$~$p$ and $P(1)$~$=$~$1-p$.
It turns out that sign fluctuations strongly affect the nature of the ground state.
No matter what is the initial state, a Dirac cone state or a spinon Fermi surface state, as the probability of sign flips reaches a threshold $p$~$=$~$17\%$, an almost flat band with large density of states appears at energy around zero, as shown in the inset of Fig.~\ref{kn4}(c). The original flat band(s) of the unperturbed kagome lattice locating at the top of the spinon band structure disappear due to impurities. Furthermore, with such probability of sign flips,  The motion of spinons will be suppressed due to destructive interference effect.
As a result, all the spinon eigenstates are localized in the lattice space\cite{supply}.
Since the spinons are quasiparticles in strongly interacting systems, the suppression of spinon motion is essentially many-body localization.
Similar mechanism also occurs in the charge localization in doped Mott insulators \cite{2019_PRL_WengZY, 2013_scientific_WengZY}.
Within this theoretical scenario, the bulk magnetic susceptibility $\chi$ is then calculated and shown in Fig.~\ref{kn4}(c)-(d).
The low-energy (nearly) flat band resulting from the phase randomness gives rise to a prominent upturn in $\chi$ at  $T\le{0.101}~J$ and a peak at about  $T$~$=$~$0.0063~J$, followed by a drop at even lower temperatures.

The above theoretical results of $\chi$ are in good consistency with the experimental data of $K_{\rm{n}}$ in respect to their overall shape and their characteristic temperatures. Taking $J$~$\approx$~79~ K~\cite{2022_PRB_SLLi2}, then the shoulder appears at around $0.51J\sim$~40~K and the upturn starts at around $0.101J\sim$~8~K, which are in perfect agreement with the Knight shift data in Fig.~\ref{kn4}(b). At the extremely low temperature region, a peak shows up at 0.5~K, which is slightly lower than that of $K_{\rm{n}}$. The appearance of the peak is due to the large density of states of the almost flat band with very low but nonzero energy (see the inset of Fig.~\ref{kn4}(c)).
We calculated the fraction of the integrated weight of spinons in the flat band, and found that 6.3\% spinons are sufficient to account for the observed upturn in the calculated $\chi$.
The peaked behavior, observed in $K_{\rm n}$, shown at temperatures below 1.5~K, are consistent with flat band picture.
This is in contrast to a Curie-Weiss behavior as expected for magnetic impurity effect.


Therefore, the large resemblance in the temperature dependence of the experimental data (Fig.~\ref{kn4}(a)) and the theoretical results (Fig.~\ref{kn4}(c)), in a rather large temperature range, supports that our measured $K_{\rm{n}}$ is strongly affected by disordered fermionic excitations rather than 3D couplings.
Besides, the shouldered behavior in the specific heat data, calculated from the flat band around Fermi surface~\cite{supply}, also agrees well with experiment~\cite{2022_PRB_SLLi2}.


\section{\label{sslr}Spin-lattice relaxation rate }

\begin{figure}[t]
\includegraphics[width=8.5cm]{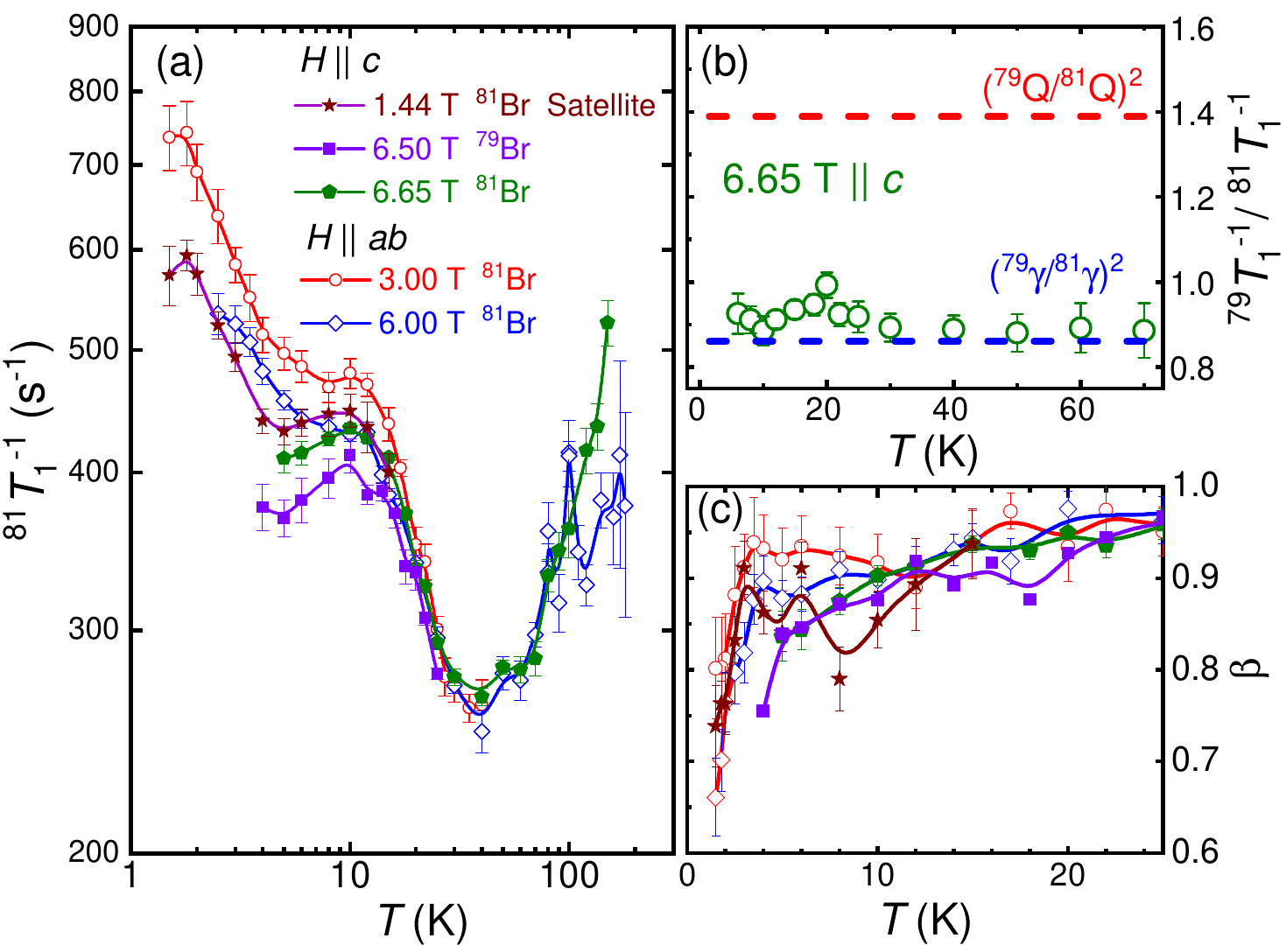}
\caption{\label{invT15}
\textbf{Spin-lattice relaxation rate.}
(a) 1$/T_1$ as functions of temperatures, measured on the center and satellite lines of $^{81}$Br(1) at different fields.
(b) Ratios of 1$/T_{1}$ on two types of isotopes, measured at 6.65~T.
The horizontal lines mark the values of ($^{79}$Q/$^{81}$Q)$^2$ and ($^{79}\gamma$/$^{81}\gamma$)$^2$, respectively.
(c) Stretch factors $\beta$ for fitting the spin-recovery curve.
}
\end{figure}

Spin-lattice relaxation rate $1/T_{1}$ is a sensitive probe of low-energy fluctuations.
In Fig.~\ref{invT15}(a), $1/T_1$ for $^{81}$Br(1) are shown as functions of temperatures.
$1/T_1$ increases with temperature from 40~K to 200~K, which should be a signature of AFM fluctuations.
Data at typical field below 3~K are not presented, where the stretching factor $\beta\le$~0.6 and therefore $1/T_1$ data are not reliable.
Below 1.5~K, the longitudinal spin recovery (data not shown) exhibits a two-exponent behavior for
unclear reasons, which prevents us to make further studies on $1/T_1$  at lower temperature.

For nuclei with quadruple moments, $1/T_{1}$ is contributed by both magnetic fluctuations and lattice fluctuations through couplings to the local hyperfine fields and EFGs, respectively.
Thanks to different Zeeman factors and quadrupole moments of $^{79}$Br and $^{81}$Br, these two contributions can be identified, as the spin fluctuation part is proportional to $\gamma_n^2$ and the lattice fluctuation part is proportional to $Q^2$.
For example, the spin contribution to $1/T_1$ is written as $1/T_1$~$=$~$\gamma_n^2 k_{\rm{B}} T/\mu_{\rm{B}}^{2}\sum _{q}A_{\rm{hf}}^2(q)$Im$\frac{\chi(q,\omega)}{\omega}$, where $\chi(q,\omega)$ is the dynamical susceptibility, $\omega$ is the Larmor frequency of the nuclei and $A_{\rm{hf}}$ is the hyperfine coupling constant.

In Fig.~\ref{invT15}(b), the ratio of 1$/^{79}T_1$ and 1$/^{81}T_1$ is plotted as a function of temperatures.
Two horizontal lines, with constant values of ($^{79} \gamma /^{81}\gamma$)$^2$~$=$~0.861 and ($^{79} Q /^{81}Q$)$^2$~$=$~1.389, are added as references which set the limit for pure magnetic and pure structural fluctuation cases, respectively.
For temperatures from 8~K to 70~K, $^{79}T_1^{-1}/^{81}T_1^{-1}$ falls close to the lower line, which indicates that magnetic fluctuations dominate in $1/T_1$ at temperatures below 70~K, whereas no structural fluctuations are found. Therefore, we can conclude that there no structural instability from 70~K down to 5~K.

With temperatures below 30~K, $1/T_1$ shows a prominent upturn upon cooling, which suggests the development of low-energy spin fluctuations.
 Furthermore, $1/T_1$ does not change with field values and orientations until below 15~K, which suggests that the spin fluctuations at temperatures between 15~K and 30~K are very isotropic.
The large onset temperature of the spin fluctuations and its isotropic behavior may not be described by the 3D AFM fluctuations induced by interlayer couplings, for the following two reasons.
First, the 3D AFM fluctuations in the lattice is usually anisotropic. Second, the energy scale of the interlayer coupling is about 2.4~K~\cite{2022_PRB_YSLi,2020_JMMM_JXM}, which is too small to account for such a high onset
temperature of upturn. In particular, the antisite disorder on the Br(2) sites,
which sets in between the Cu$^{2+}$ layers, may further reduce the 3D couplings.
On the contrary, non-magnetic impurities, as described in Section~\ref{skn}, may
lead to such enhancement of isotropic spin fluctuation at $T{\le}J/2$,
as observed by the temperature-dependence of $K_{\rm{n}}$ and
$\chi$ (Fig.~\ref{kn4}(a)-(d)).

Below 8~K, however, $1/T_1$ exhibits an upturn with $H||ab$, but a downtown with $H||c$,
and $1/T_1$ is also reduced with field in both orientations below 1.8~K.
The dramatic anisotropic and the field-suppression effects  in $1/T_1$ may result from the development of anisotropic AFM fluctuations due to interlayer couplings and spin-orbit coupling such as the Dzyaloshinskii-Moriya (DM) interactions,
with easy-axis along the $c$ axis.
1$/T_1$ increases when cooled from 5~K to 1.8~K with an upturn at low temperature, which in principle should indicate gapless excitations
which will be further discussed with the 1$/T_2$ data.
In fact, the linewidth of Br(1) spectra, as shown in Fig.~\ref{spec3}(c), also increases
dramatically below 8~K, which are consistent with the development of short-range correlations,
though the ordering tendency is suppressed by the inherent antisite disorder in the system.
In this temperature range, disorder-induced spin fluctuations, though prominent as shown in $\chi$ (Fig.~\ref{kn4}(b)), may overlap with
the 3D anisotropic fluctuations as discussed above.
The $1/T_2$, on the other hand, reveals dominate isotropic spin fluctuations as shown in the following section.

\section{\label{sssr}Spin-spin relaxation rate}

\begin{figure}[t]
\includegraphics[width=7.5cm]{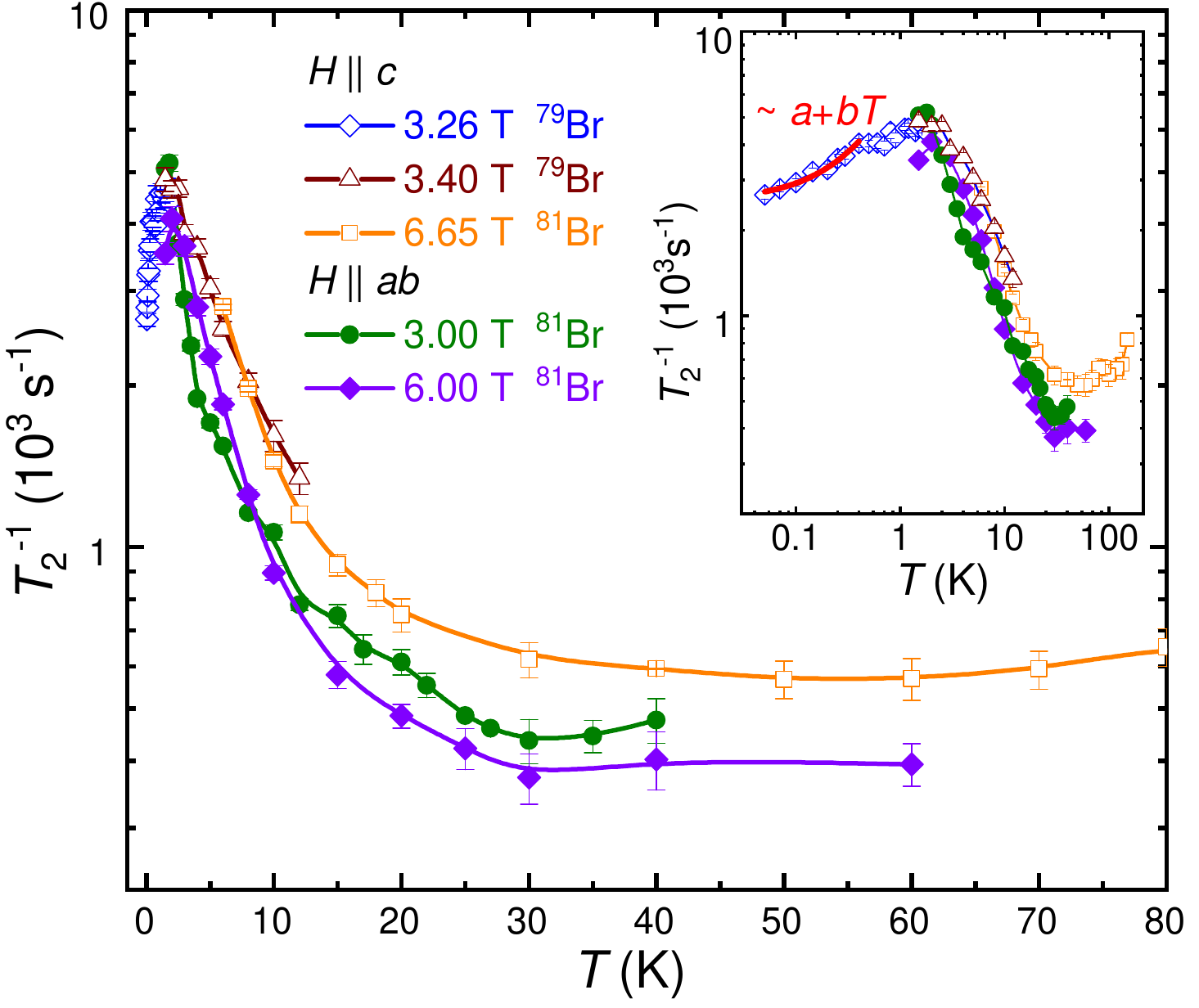}
\caption{\label{invT26}
\textbf{Spin-spin relaxation rate.}
$1/T_2$ as functions of temperatures measured with different fields.
Inset: log-log plot of $1/T_2$. The solid line represents a
linear fit of $a$$+$$bT$ to the low-temperature data.
}
\end{figure}

To further study the low-energy spin dynamics, the spin-spin relaxation rate $1/T_2$ was measured.
The detailed data with different field values and orientations are shown in Fig.~\ref{invT26}(a).
Compared to the $1/T_1$ data, the overall trend of $1/T_2$ with temperature seems to be similar, where a dip forms at about 30~K, followed by an upturn upon cooling.
However, there are also distinctive behaviors in the $1/T_2$.

First, the stretching factor $\beta$ remains as one in the whole temperature range, by fitting transverse spin recovery to obtain $T_2$, as described in Section~\ref{Materials}.
Therefore, $1/T_2$ reveals intrinsic behavior of the compound.

Second, from 30~K to 2~K, no kinked behavior is found in the log-log plot of $1/T_2(T)$ (Fig.~\ref{invT26} inset). In particular,
at temperatures below 8~K, the changes of $1/T_2$ with field orientations
and field amplitudes are very small, which suggests that $1/T_2$ is dominated by
isotropic spin fluctuations in the whole temperature range, compared to the
anisotropic fluctuations in $1/T_1$ below 8~K.
This suggests that dominant fluctuation revealed by $1/T_2$ should be affected by another mechanism rather than 3D coupling effects.
In fact, the scattered spinons, as discussed in Sec.~\ref{skn}, can account for such upturn in a large temperature range below $T{\le}J/2$,
by enhanced low-energy spin fluctuations from the flat band, from the spinon Hamiltonian with phase randomness discussed in Sec.~\ref{skn}.

Lastly, a remarkable peak develops in $1/T_2$ at about 1.5~K, below which $1/T_2$ decreases dramatically.
In this aspect, $1/T_2$ behaves very similar to $K_{\rm{n}}$ by showing a peaked behavior, albeit at a lightly higher temperature.
These observations are all aligned with the flat band of scattered spinons at a finite energy of  $0.0063~J$ (Sec.~\ref{skn}).
The different peak temperatures in $K_{\rm{n}}$ and $1/T_2$ may be related to a shorter time scale for measurements of $1/T_2$.

We attempted to fit $1/T_{2}$ by the form $1/T_2$~$=$~$a+bT$, at temperatures between 50~mK and 400~mK. As shown by the solid line in the inset of Fig.~\ref{invT26}, the success of the fitting
suggests additive contributions to $1/T_2$: a constant term and a linear-$T$ term.
In principle the linear-$T$ term presents gapless excitations with a Fermi surface, which may be caused by a Fermi surface-like behavior at high temperatures, by picking up the peaked spectra at finite energies as shown in Fig.~\ref{kn4}(c) (inset).

\section{Discussions}

In YCu$_3$-Br, the behavior of $K_{\rm{n}}$, $1/T_1$ and $1/T_2$ at all temperatures can be accounted by antisite randomness, 3D spin fluctuations (resulting from interlayer coupling) and DM interactions.
Note that only $1/T_1$ shows anisotropic spin fluctuations which should be caused by the interlayer couplings and DM interactions.
We also note that the external field suppresses the anisotropic short-ranged AFM correlations as revealed by $1/T_1$, which may help to stabilize a QSL ground state as revealed by a recent work~\cite{2023_arxiv_LuLi}, and deserves further study.

Our study suggests that non-magnetic antisite randomness seems to be crucial in 2D KHAF.
On one hand, DM interactions, which give rise to AFM long range-order in YCu$_3$(OH)$_6$Cl$_3$~\cite{2019_PRB_JXMi,2019_PRB_Zorko,2019_PRM_Fabrice}, are suppressed by strong disorder in YCu$_3$-Br.
On the other hand, the antisite disorder between Br(2) and OH$^-$ and Y atom displacement, may cause bond randomness, namely, fluctuating exchange couplings in the Cu-O-Cu path~\cite{2022_PRB_YSLi} within a finite energy range around $J$.
In our mean-field model, this fluctuations in exchange couplings may change the ground state of this system.



At the mean-field level, the randomness in the bond interactions can cause both amplitude fluctuations and phase fluctuations of the spinon kinetic terms. If the phase randomness $p$ reaches 17$\%$, on top of a Fermi surface like density of states, the spinon dispersion has an almost flat band close to the Fermi energy.
The physical responses of the low-energy flat band agree well with the experimental data, including the linear temperature-dependence in $1/T_2$ and the sharp peak in $K_{\rm{n}}$ at 0.5~K.
The nearly flat energy band corresponds to low-energy spinons excitations which are spatially localized.
The existence of large density of states close to zero energy is quite different from the picture of random singlet states~\cite{1994_PRB_Fisher,2010_PRL_Singh}.

Interestingly, besides the spinons in the nearly flat band, almost all of the spinon eigenstates are localized in lattice space, as shown in the SM~\cite{supply}. Since the spinons are collective excitations resulting from strong interactions, the localization of spinons are essentially many-body localization caused by the randomness in the interaction strength. The localized spinons contribute to both specific heat and susceptibility, but not to the thermal transport. As a by-product, 
our work indicates that the thermal conductance of YCu$_3$-Br contributed by the spinons is very small~\cite{2022_PRB_Hess}. Localized magnons are not considered 
since there is no evidence of long-range magnetic order in the system.

It is enlightening to compare the localized spinons with isolated magnetic impurities despite that our sample is of high quality. Here we list several essential differences. Firstly, the density of impurities is generally very small, but the localized spinons have a flat band at low energy which has a very large density of states. Secondly, the localization of spinons results from strong spin-spin interactions, and as a result the localized spinons interact with each other via internal gauge fields (hence has stronger quantum fluctuations). On the other hand, the isolated magnetic impurities weakly  interact with each other. Thirdly, the localized spinons obey fractional (fermionic) statistics while the energy of isolated impurities obeys Bose distribution (or approximately Boltzmann distribution since the density is very low). Consequently, the magnetic susceptibility (and Knight shift) of the latter obey the Curie law but that of the former deviates from the  Curie law due to the fermionic statistics.
For simplicity, we didn't include possible DM interactions in the theoretical simulations.

\section{Summary}

In summary, our study provides spectroscopic evidence for the absence of AFM ordering in YCu$_3$-Br with temperatures down to 50 mK.
The $K_{\rm{n}}$, $1/T_1$, and $1/T_2$ data at temperatures above 30~K reveal intrinsic behavior of the 2D KHAF, unaffected by disorder.
The effect of the site-randomness disorder  leads to enhanced low-energy spin fluctuations below 30~K, consistent with our mean-field calculation of magnetic susceptibility,
by considering amplitude and sign randomness in the spinon kinetic term which give rise to a Fermi-surface like band structure.
In particular, both $K_{\rm{n}}$ and $1/T_2$ are peaked at very low temperatures, which are
consistent with the peak in the spinon density of states contributed from the nearly flat band close to zero energy.
Our results suggest that non-magnetic site disorder has a significant impact on the spinon dispersion in the $S$~$=$~$1/2$ KHA system and may cause many-body localization.

\section{ Acknowledgements}

We would like to thank Prof. Rong Yu and Prof. Wei Zhu for helpful discussions.
This work was supported by the National Key R$\&$D Program of China (Grant Nos.~2023YFA1406500,
2022YFA1402700, 2022YFA1403402, 2022YFA1405300 and 2021YFA1400401),
the National Natural Science Foundation of China (Grant Nos.~12134020,  12374156, 12374166 and 12104503),
and the Strategic Priority Research Program(B) of the Chinese Academy of Sciences (Grant No.~XDB33010100).


%

\end{document}